\documentstyle[12pt]{article}

\begin{document}

\begin{titlepage}

\begin{center}
{\Large {\bf A Group of Galaxies at Redshift 2.38$^1$}}
\end{center}

\medskip
Accepted for Publication in Astrophysical Journal: To Appear Feb 1st 1996.

\medskip

\begin{center}

Paul J. Francis$^2$
\\
{\it University of Melbourne, School of Physics, Parkville, Victoria 3052,
Australia. e-mail:~pjf@physics.unimelb.edu.au}
\\
Bruce E. Woodgate$^2$ \\
{\it NASA Goddard Space-Flight Center, Code 681,
Greenbelt, MD 20771. e-mail:~woodgate@uit.dnet.nasa.gov}
\\
Stephen J. Warren \\
{\it Imperial College, Astrophysics Group, Blackett
Laboratory, Prince Consort Road, London SW7 2BZ, UK.
e-mail:~s.j.warren@ic.ac.uk}
\\
Palle M{\o}ller \\
{\it Space Telescope Science Institute, Baltimore MD 21218,
on assignment from the Space Science Department of ESA.
e-mail:~moller@stsci.edu}
\\
Margaret Mazzolini \\
{\it University of Melbourne, School of Physics, Parkville, Victoria 3052,
Australia. e-mail:~marg@physics.unimelb.edu.au}
\\
Andrew J. Bunker \\
{\it University of Oxford, Department of Physics,
Keble Road, Oxford
OX1~3RH, UK. e-mail:~a.bunker1@physics.oxford.ac.uk}
\\
James D. Lowenthal$^3$ \\
{\it Lick Observatory, Santa Cruz CA 95064.
e-mail:~james@lick.ucsc.edu}
\\
Ted B. Williams \\
{\it Rutgers University, Physics and Astronomy
Department,
P. O. Box 849, Piscataway, NJ 08855-0849.
e-mail:~williams@fenway.rutgers.edu}
\\
Takeo Minezaki \\
{\it Department of Astronomy, Faculty of Science,
University
of Tokyo, Bunkyo-ku, Tokyo 113, Japan.
e-mail:~umineza@c1.mtk.nao.ac.jp}
\\
Yukiyasu Kobayashi \\
{\it National Astronomical Observatory, Mitaka,
Tokyo 181, Japan. e-mail:~yuki@merope.mtk.nao.ac.jp}
\\
\& \\
Yuzuru Yoshii \\
{\it Institute of Astronomy, University of Tokyo, Mitaka,
Tokyo 181, Japan. e-mail:~yoshii@omega.mtk.ioa.s.u-tokyo.ac.jp}

\end{center}

\medskip

\vspace{2mm} \noindent
$^1$Based on observations collected at Cerro Tololo Interamerican
Observatory, the European Southern Observatory (La Silla, Chile), Siding
Spring Observatory, the Australia Telescope, and the Anglo-Australian
Observatory

\vspace{2mm} \noindent
$^2$Visiting Astronomer, Cerro Tololo Interamerican
Observatory. CTIO is operated by AURA, Inc.\ under contract to the
National Science Foundation.

\vspace{2mm} \noindent
$^3$Hubble Fellow

\end{titlepage}

\begin{abstract}

  We report the discovery of a group of galaxies at redshift $2.38$. We
imaged $\sim 10$\% of a claimed supercluster of QSO absorption-lines
at $z=2.38$ (Francis \& Hewett 1993). In this small field (2$^{\prime}$ \
radius) we detect two Ly-$\alpha$ emitting galaxies. The discovery of two
such galaxies in our tiny field supports Francis \& Hewett's interpretation
of the absorption-line supercluster as a high redshift `Great Wall'.

One of the Ly-$\alpha$ galaxies lies $22^{\prime \prime}$ from a background
QSO, and may be associated with a multi-component Ly-$\alpha$ absorption
complex seen in the QSO spectrum. This galaxy has an extended ($\sim 50$kpc)
lumpy Ly-$\alpha$ morphology, surrounding a compact IR-bright
nucleus. The nucleus shows a pronounced break in its optical-UV colors at
$\sim 4000$\AA\ (rest-frame), consistent with a stellar population of
mass $\sim 7 \times
10^{11} {\rm M}_{{\rm Sun}} $, an age of $> 500$Myr, and little on-going
star-formation. C~IV emission is detected, suggesting that a concealed
AGN is present. The Ly-$\alpha$ emission is redshifted by $\sim 490
{\rm km \ s}^{-1}$ with respect to the C~IV emission, probably due to
absorption. Extended H-$\alpha$ emission is also detected; the ratio of
Ly-$\alpha$ flux to H-$\alpha$ is abnormally low ($\sim 0.7$), probable
evidence for extended dust.

  This galaxy is surrounded by a number of very red ($B-K>5$) objects,
some of which have colors suggesting that they too are at $z=2.38$. We
hypothesize that this galaxy, its neighbors and a surrounding lumpy gas
cloud may be a giant elliptical galaxy in the act of bottom-up formation.

Note: The figures for this paper are available in this archive, by
anonymous FTP to tauon.ph.unimelb.edu.au in the directory
/incoming/pjf/paper, or on the WWW from
\begin{verbatim}
http://www.ph.unimelb.edu.au/~pjf/blob.html
\end{verbatim}

\end{abstract}

\section{Introduction}

The extent of galaxy clustering at high redshifts ($z>1$) is controversial.
CDM and similar models predict that most large-scale structure forms at
redshifts below one. This is consistent with the rapid evolution
in galaxy clusters seen between $z \sim 0.5$ and the present (
Butcher \& Oemler 1978, Edge et al. 1990). Further
evidence comes from the apparently weak clustering of faint blue
galaxies at $z \sim 0.7$ (Efstathiou et al. 1991) and the
weak clustering of Ly-$\alpha$ forest clouds (eg. Carswell
\& Rees 1987).

On the other hand, an increasing body of data suggests that galaxies
at redshifts above one are strongly clustered on large scales. Evidence
comes from metal-line QSO absorption systems (eg. Heisler, Hogan \& White 1989,
Jakobsen \& Perryman 1992, Foltz et al.
1993, M{\o}ller 1995, Williger et al. 1995), from
the association of Ly-$\alpha$ emitting galaxies with damped Ly-$\alpha$
absorption systems and QSOs (Wolfe 1993, Djorgovski et
al. 1985), and from the often serendipitous discovery of individual
clusters (eg. Dressler et al. 1993, and possibly
Giavalisco, Steidel \& Szalay 1994).

Francis \& Hewett (1993)
claim to have discovered two coherent structures of gas, at least
ten co-moving Mpc in size, one at redshift 2.38, the other at 2.85.
These postulated structures cause Ly-$\alpha$ absorption at matching
redshifts in the spectra of two background QSOs separated by 8$^{\prime}$ .
The authors discuss two possibilities: either the gas structures are
primordial pancakes, or `Great-Wall' style galaxy superclusters.
In both cases, these structures would be the most dramatic examples yet of
large scale structure at high redshift.

In this paper, we attempt to confirm the existence of one of Francis \&
Hewett's enormous structures. We have studied a small part of the
postulated z=2.38 structure, using a wide range of optical, infra-red
and radio observations on ten different southern hemisphere telescopes (\S~2).
Our initial aim was to search for Ly-$\alpha$ emitting sources at this
redshift, using narrow-band Fabry-Perot imaging. We found at least two
Ly-$\alpha$ emitting galaxies at z=2.38 (\S~3.1) one of which has a
remarkable extended Ly-$\alpha$ morphology. Extensive follow-up
observations at many frequencies revealed a number of other possible
galaxies at z=2.38.

We describe our observations in \S~2, results are shown in
\S~3, and discussed in \S~4. Throughout this paper, we will assume
that $H_0 = 75\ {\rm km\ s}^{-1}{\rm Mpc}^{-1}$ and $q_0 = 0.5$.

\section{Observations}

Due to limited observing time and poor weather, we were able to observe only
a small part of one of Francis \& Hewett's postulated structures. We
chose to study the $z=2.38$ structure, and imaged the field around
one of the two background QSOs, 2139-4434. The absorption line system
in 2139-4434 at $z=2.38$ consists of at least three components, a central
one of wavelength 4108.5\AA\  and neutral hydrogen column density $\sim 7
\times 10^{18} {\rm cm}^{-2}$, and two lower column density systems
(Fig~4). Given the limited
resolution of our spectra, we have not attempted a decomposition of the
two subsidiary absorption systems, but we estimate their wavelengths
as being 4101\AA \ and 4119 \AA . All three may well be blends of lower
column-density lines.

\subsection{Optical Imaging}

The first step in our imaging campaign was to search for redshifted
Ly-$\alpha$ emission from galaxies at $z=2.38$.
A field surrounding the QSO 2139-4434 was imaged on the
nights of 1994 June 4th and June 5th, using the
Rutgers Fabry-Perot system (Gebhardt et al. 1994)
with Goddard etalons, on the CTIO 4-m
telescope. The etalon was set to a central wavelength of 4110 \AA \
with a width of 30 \AA . The relatively large width was chosen to allow
for possible peculiar velocities relative to the main absorption system,
as suggested by the wavelengths of the subsidiary absorption systems.
We alternated 1800 second exposures on-band with 360 second exposures
off-band. Off-band exposures were also taken through the etalon, but with
the plate spacing altered to shift the central wavelength away from the
absorption line wavelength, and with
the order-blocking filter removed. This increased the count rate from
continuum sources by a factor of $\sim 5$. The similarity of the set-up
between on- and off-band exposures was intended to prevent any ghost
images showing up as narrow-band excess objects.
The telescope was offset by $\sim 5^{\prime \prime}$ \ in a random direction
between pairs of on- and
off-band exposures to improve flat fielding and as an additional
precaution against ghosts.

Six pairs of on- and off-band images were taken on each night (Fig~1). Seeing
was $\sim 1.6^{\prime \prime}$ \ (Full Width at Half Maximum Height, FWHM)
throughout.
Images taken on the first night
suffered from poor throughput in the order-blocking filter; a more
efficient filter was used on the second night. A narrow-band flux limit of
$8.4 \times 10^{-17} {\rm \ erg\ cm}^{-2}{\rm \ s}^{-1}$ for a point
source was reached ($3 \sigma$).
All images were reduced using standard
IRAF\footnote{IRAF is distributed by the National Optical
Astronomy Observatories,
which is operated by the Association of Universities for Research in
Astronomy, Inc. (AURA) under cooperative agreement with the National
Science Foundation.} procedures, then shifted and added using inverse
variance weighting to maximize the final signal-to-noise ratio. Due to
light cirrus during the observations, only relative photometry was obtained,
bootstrapped from CCD-calibrated UK Schmidt plate scans. This introduces
an uncertainty of $\sim 0.15$ magnitudes in the photometry.

Candidate narrow-band excess objects were selected in a variety of ways.
On- and off-band images were blinked and compared by eye, and both aperture
photometry and DAOPHOT were used. All approaches were cross checked and
found to give consistent answers. Automatic procedures needed substantial
human intervention, due to the strong signal-to-noise gradient across
the frames caused by the shifting and adding, which principally manifested
itself in spurious sources near the edge of the field. All fluxes quoted
in the paper are total fluxes. For fainter unresolved objects, photometry
was carried out using a small aperture (diameter $\sim 1.6^{\prime \prime}$,
the seeing FWHM) and
corrected up to a total flux using the PSF of bright isolated stars in
the field. Flux and magnitude limits are quoted for the area around QSO
2139$-$4434 where the noise is a minimum, and for an aperture of diameter
$\sim 1.6^{\prime \prime}$.

An additional $B$-band image was obtained on the CTIO 0.9-m using the Tek 2k
chip, giving a $12^{\prime} \times 12^{\prime}$ field centered on a point
mid-way between the two background QSOs
and including the fields around both QSOs. A total
of 10800 seconds exposure was obtained on the night of 1994 June 7th.

$I$ and $B$ band images of the field around 2139-4434 were obtained on the
night of 1994 August 5th with the EFOSC camera (Eckert, Hofstadt
\& Melnick 1989) on the ESO 3.6-m telescope.
Exposure times were 2700 s in $B$ and 1620 s in $I$, and the seeing was poor
($\sim 2.5^{\prime \prime}$ ). The $I$-band image reached a limiting
magnitude of $23$
($3 \sigma)$.
A weighted sum of the two $B$-band images reached a limiting magnitude of
$24.7$ ($3 \sigma$). Photometry of these frames was carried out as above.

Astrometry for all frames was bootstrapped from positions in the COSMOS
database, maintained on-line at the Anglo-Australian Observatory. In each
image, 5 stars were used to obtain a 6-coefficient plate solution. Residuals
from the fit (rms) in each coordinate were $0.15^{\prime \prime}$ for the CTIO
4-m images, $0.12^{\prime \prime} $ for the ESO 3.6-m images, and
$0.12^{\prime \prime}$ for the
Siding Spring 2.3-m IR-images.

\subsection{IR Imaging}

IR imaging was obtained on the nights of 1994 September 25th to September
28th on the Siding Spring Observatory 2.3-m telescope, using the PICNIC
near-IR camera (Kobayashi et al. 1994). The detector
field-of-view was centered
midway between the QSO and galaxy B1 (see \S~3.1). PICNIC has a
$256 \times 256$  NICMOS chip, with  $0.5^{\prime \prime}$ pixels.
250 exposures, each of 17 seconds, were obtained in both $H$ and
$K^{\prime}$ bands, rastered on a $5 \times 5$ grid, in $1.0^{\prime \prime}$ \
seeing. Limiting magnitudes were 21.0 in both $H$ and $K^{\prime}$
(3 $\sigma$). 75 exposures, each of 67 seconds, were taken through an
interference filter, of central wavelength $2.238$  $\mu$m \ and width
(FWHM) $0.048$  $\mu$m , roughly matching the expected wavelength of
H-$\alpha$ emission at $z=2.38$. The $K^{\prime}$ image was used as the
off-band for this H-$\alpha$ image, and a ($3 \sigma$) flux limit of
$4.5 \times 10^{-16}{\rm \ ergs\ cm}^{-2}{\rm \ s}^{-1}$ was reached.
Conditions
were photometric, accuracy being limited by a slight time-dependence in
the flat field to $\sim 0.1$ magnitudes.

Further IR imaging was obtained with the IRAC-2B camera on the ESO/MPI 2.2-m.
22 exposures of 300 seconds were taken centered on the QSO in the J band,
in non-photometric conditions on 1994 Oct 25th. Photometric calibration
was obtained from 5 exposures of 180 seconds each taken on 1994 November
18th, with the same instrument. Seeing was 1.1$^{\prime \prime}$ \ and the
$3 \sigma$ magnitude limit is $21.77$. 29 exposures of 300 seconds were
also taken in $K^{\prime}$ in photometric conditions on 1994 Oct 24th.

\subsection{Spectroscopy}

We detected three possible Ly-$\alpha$ emitting galaxies in our
narrow-band imaging, B1, B2 and B3 (see \S~3.1 below). Follow-up
spectroscopy of two of these candidates, B1 and B2, was
obtained on the nights of 1994 Aug 4th and 5th with the EMMI
instrument (Dekker et al. 1991) on the
NTT. The slit was placed over the background QSO at such an angle that
it would also cover the galaxy candidate. In the case of B1, the slit
passed through the peak narrow-band emission, but 2$^{\prime \prime}$ \
south-west
of the IR-bright nucleus. Blue arm spectra of both objects were
taken with BLMD, with the \# 3 grism and a 2$^{\prime \prime}$ \ slit giving a
resolution of $2.35$\AA .
Exposure times were 7200 seconds for both candidates. In addition, a red
arm spectrum of B1 was taken with RILD, grating \# 5, a
1.5$^{\prime \prime}$ \ slit and an
exposure time of 4800 seconds, the resolution being $\sim 7$\AA .

Additional spectra of B1 were taken at the AAT on September 6th 1994.
The slit was laid along the major axis of B1 and 9000 seconds integration
obtained. Seeing was $\sim 4^{\prime \prime}$ , preventing us from obtaining
any useful spatial information. The RGO spectrograph, with the 600V
grating set blaze to collimator, and the faint object red spectrograph,
were used simultaneously, with the dichroic.

\subsection{Radio Observations}

The field containing all the galaxy candidates was observed with the
Compact Array of the
Australia Telescope on Jan 26th and Jan 28th 1995. A total of 20 hours
of observations were obtained both at 1344 and 2378 MHz, with the 6A
configuration. The data were reduced using standard procedures embodied in the
MIRIAD package, and CLEANed with a restoring beam of $6.8 \times 4.7$
arcsec at 1344~MHz and $3.8 \times 2.5$ arcsec at 2378 MHz.
The field has also been surveyed at 847 MHz
with the Molonglo Synthesis Telescope (Hunstead, private communication).
Although radio-quiet, QSO 2139-4434 is detected at all three frequencies,
with fluxes of $3.8 \pm 1.1$ mJy at 847 MHz, $1.61 \pm 0.09$ mJy at 1344 MHz
and $0.86 \pm 0.09$ mJy at 2378 MHz.
Nothing else is detected in the field
at any wavelength, $3 \sigma$ limits being $3.3$ mJy (847 MHz) and
$0.27$ mJy (1344 and 2378 MHz).

\section{Results}

\subsection{Line-Emitting Galaxies}

We detected three candidate Ly-$\alpha$ emitting galaxies, B1, B2 and B3
(Fig~1, Fig~2). All three were seen in the narrow-band Ly-$\alpha$
image, but not in the off-band Fabry-Perot image, though B3 is only
detected with $\sim
2.7 \sigma$ confidence. Our spectra of B1 and B2 show that the
narrow-band image flux is indeed due to a strong isolated emission-line.
For B1 we also detect C~IV and H-$\alpha$ emission, confirming the
identification of the line as Lyman-$\alpha$. The properties of these
candidates are shown in Table~1.

Object B1 lies 22$^{\prime \prime}$ \ from the background QSO, and has an
extended
emission-line morphology (Fig~3). A knot of Ly-$\alpha$ flux lies at
the Southwest
of B1, marginally resolved (FWHM $\sim 4^{\prime \prime} $, 18kpc) and
contributing
roughly half the flux. The remainder comes from a tail extending $\sim
10^{\prime \prime}$ \ (50 kpc) to the North-East.

Object B1 has an H-$\alpha$ flux of $(9.2 \pm 1.5) \times 10^{-16}{\rm \ erg\
cm}^{-2}{\rm \ s}^{-1}$ ($1 \sigma$ errors). The H-$\alpha$ emission
morphology is similar to that of the Ly-$\alpha$ morphology. Note that
H-$\alpha$ contributes less than 10\% of the $K^{\prime}$-band flux.
C~IV is detected with $3 \sigma$ confidence, at a position on the slit
corresponding to the peak of the Ly-$\alpha$ emission. The ratio of
Ly-$\alpha$ to C~IV is $\sim $ 7:1 at this position. A possible He~II
line is seen at $\sim 2 \sigma$ confidence, about 50\% weaker than
C~IV; no other significant emission features are seen.

The redshift of B1, as deduced from the C~IV emission (Fig~5), is $2.3796$,
identical within the errors ($\sim 100 {\rm \ km\ s}^{-1}$)
to that of the centroid of the QSO absorption-line (Fig~5). However,
the Ly-$\alpha$ emission is redshifted by $\sim 490 {\rm \ km\ s}^{-1}$ with
respect to both of these lines. The Ly-$\alpha$ line is strongly red
asymmetric, and has a velocity width of $\sim 600 {\rm \ km\ s}^{-1}$. The
apparent relative displacement of Ly-$\alpha$ and C~IV may be caused by
Ly-$\alpha$ absorption (\S~4.2).

B1 is not seen in any of the broad-band optical images. No continuum is
seen in the blue spectra, though a faint continuum is seen with about
$2 \sigma$ confidence in the red spectrum, giving B1 a tentative continuum
magnitude of $25.8$ at 5000 \AA . In the near-IR however, a strong
unresolved source is seen, lying within the extended Ly-$\alpha$
morphology. This source is marginally detected in the $J$ band
($J \sim 22.0$) but is relatively bright in $H$ and $K^{\prime}$ ($H=19.9$,
$K^{\prime} = 19.0$).

The Ly-$\alpha$ knot and the IR source are misaligned by $\sim 2.5^{\prime
\prime}$ .
This misalignment is real; the two images have internal astrometry accurate
to $\sim 0.15^{\prime \prime}$ . The colors of the IR source show a pronounced
break
at $\sim 1.3\mu$m, which is the expected position of the Balmer or
$4000$\AA\ break at $z=2.38$.
Objects showing such a near-IR spectral break are rare; only four are seen
in our 7800 square arcsec field of view. Given this surface density, the
probability of such an object lying within $2.5^{\prime \prime}$ \ of the peak
of the Ly-$\alpha$ emission is $0.2$\% .
We therefore consider it highly probable that the IR source is at the
same redshift as, and associated with, the Ly-$\alpha$ emission.
H-$\alpha$ emission comes from the same region as Ly-$\alpha$ emission;
the H-$\alpha$ emission is detected from both the Ly-$\alpha$ knot and
tail with $> 4 \sigma$ confidence, having approximately equal H-$\alpha$
fluxes. The H-$\alpha$ image also shows flux at the position of the
continuum IR source, no more that would be expected from a featureless
continuum spectrum. The signal-to-noise ratio of the H-$\alpha$ image is
too poor to allow any other morphological information to be extracted.

Object B2 lies 63$^{\prime \prime}$ \ from the background QSO, and is
unresolved. It is
marginally detected ($\sim 2 \sigma$ confidence) in the $K^{\prime}$
image ($K^{\prime} \sim 20.8$), and not detected in any other band. We
only detect the one line in B2, but its equivalent width is much greater
than that of [O~II] in a typical field galaxy (Colless et al.
1990, Glazebrook et al. 1995), strongly suggesting that the line
is indeed Ly-$\alpha$ at $z=2.38$. H-$\alpha$ is not detected, the upper
limit being $4.5 \times 10^{-16} {\rm \ erg\ cm}^{-2}{\rm \ s}^{-1}$ ($3
\sigma$).
B2's Lyman-$\alpha$ emission is redshifted by $160 {\rm \ km\ s}^{-1}$ with
respect to the absorption-line centroid, has a red asymmetric profile, and
a velocity width (FWHM) of $\sim 450 {\rm \ km\ s}^{-1}$. No other lines
are seen in the blue spectrum, the
upper limit on their flux being 30\% that of B2, but the spectrum
only covers the rest-frame wavelength region 1182--1320 \AA , so no other
strong lines would be expected.

B3 is detected with only $\sim 2.7 \sigma$ confidence; we have no color or
spectral information on it.

\subsection{Broad-band Colors}

When we obtained IR images of our field, we were surprised to find that
the QSO 2139-4434 and B1 are surrounded by a group of very red
($B-K^{\prime} > 5$) objects, extending over $\sim 50^{\prime \prime}$ \
(Fig~6).
Multicolor photometry gives us some clues to the redshifts of these
objects.

B1 itself, as discussed above, shows a clear spectral break at about $1.3
\mu$m (Fig~7), which we ascribe to the redshifted Balmer or $4000$\AA\ break,
giving it an
$I-H$ color of $3.1$. Most of the other red objects near it
have $I-H < 2.5$; they have relatively flat spectral energy distributions
between the $I$ and $K^{\prime}$ bands, and very red $B-I$ colors. Spectral
energy distributions of the two brightest of these objects are shown in
Fig~7 as examples. These
objects are unlikely to be associated with the Francis \& Hewett supercluster;
instead their colors are typical of galaxies at $0.3 < z < 1.2$.

A few of B1's red neighbors do have colors consistent with a redshifted
Balmer or $4000$\AA\  break at $z \sim 2.38$. These candidate supercluster
members are
marked as squares in Fig~2, and their spectral energy distributions are
shown in Fig~7. In most cases they are too faint to say with confidence
whether they show a strong redshifted spectral break; further deep $I$-band
photometry will resolve this question.

A number of other objects in the field of view but more distant from B1
also have large $I-K^{\prime}$ colors, and hence are candidate $z=2.38$
objects. These objects are marked as squares in Fig~2, and the spectral
energy distributions of the two best candidates are shown in Fig~7.

Could any of these very red objects be M-dwarfs? Leggett (1992)
showed that the optical/near-IR colors of M-dwarfs lie in a well defined
subset of the color space. B1 and the other candidate $z=2.38$ galaxies,
however, show very different colors. In particular, their $J-K^{\prime}$
colors are much redder than those of any of the dwarf stars studied by
Leggett. We therefore consider it unlikely that these objects are galactic
stars.

In conclusion, the very red objects surrounding the QSO and B1 fall into
two classes; objects with spectral breaks between the $B$ and $I$ bands,
which are probably foreground galaxies, and objects with spectral breaks
between the $I$ and $H$ bands, which are probably at the same redshift as
the QSO absorption lines, B1 and B2.

\subsection{The QSO Sight-line}

The line of sight to the background QSO, 2139$-$4434, lies 22$^{\prime \prime}$
\ from B1.
We performed a point spread function (PSF) subtraction on our images to look
for a possible absorbing galaxy closer to the line of sight.

The PSF subtraction in the Ly-$\alpha$ and $B$-band images revealed nothing.
The QSO is not black in the Ly-$\alpha$ image as our filter passband was
wider than the absorption line. We would not detect objects within
$\sim 1^{\prime \prime}$ \ of the QSO sight line.

In the $I$ band and the IR images, a source is seen $1.8^{\prime \prime}$ \
from the
QSO sightline, lying roughly $30^{\circ}$ east of north. This source is
most clearly seen in the $K^{\prime}$ image.
This object has a flat spectral energy distribution between $I$
and $K^{\prime}$ but is very red in $B-I$ ($B > 24$, $I  \sim 20.4$,
$K^{\prime}
\sim 18.3$). These colors are very similar to those of the candidate
foreground galaxies discussed in \S~3.2 above, and we therefore hypothesize
that like them this object lies at a redshift $0.3 < z < 1.2$.

Note that the presence of so many intermediate redshift galaxies close to
the QSO sight-line may not be coincidental; they may be gravitationally
lensing QSO 2139-4434, amplifying it enough to bring it into the QSO sample
(eg. Webster et al. 1988, Rodrigues-Williams \& Hogan
1994).

\section{Discussion}

\subsection{The Supercluster?}

Francis \& Hewett claim that their absorption-line supercluster has an
overdensity of $> 30$ (95 \% confidence), although they were unable to
determine whether
it was some sort of primordial pancake or a `Great Wall' style supercluster.

Our detection of galaxies near one of the QSO absorption line clouds
suggests that it
is indeed a galaxy cluster and not a primordial pancake, though it may be
a very gas rich supercluster. Following the argument of Wolfe
(1993) we can estimate the overdensity of Ly-$\alpha$ emitting galaxies
in the imaged field, by comparing our galaxy detections with the null
results reported by field searches for Ly-$\alpha$ emission.

As our control sample, we will use the search for field Ly-$\alpha$
emitting galaxies carried out by Pritchet \& Hartwick (1990).
They claim to have searched a co-moving volume 100 times larger than ours,
down to a limiting flux of $5.4 \times 10^{-17}{\rm \ erg\ cm}^{-2}{\rm
\ s}^{-1}{\rm \ arcsec}^{-2}$ ($2 \sigma$). B1 has a peak surface brightness
of $1.6 \times 10^{-16}{\rm \ erg\ cm}^{-2}{\rm \ s}^{-1}{\rm \ arcsec}^{-2}$.
B2 is unresolved in $\sim 1.6^{\prime \prime}$ \ seeing; we measured a peak
surface
brightness of $9.6 \times 10^{-17}{\rm \ erg\ cm}^{-2}{\rm \ s}^{-1}{\rm
\ arcsec}^{-2}$. Both B1 and B2 thus easily exceed their quoted flux limit,
unless their seeing was very poor. Pritchet \&
Hartwick do not explicitly state their equivalent width limit, but they
imply that they were sensitive to objects showing $> 0.1$ magnitude
differences between their on- and off-band images, implying an equivalent
width limit $\ll 50$\AA . Both B1 and B2 have equivalent widths much
greater than this. It is therefore probable that if any
objects such as B1 and B2 existed in the region surveyed by Pritchet \&
Hartwick, they would have been found.

We calculate limits on the overdensity of Ly-$\alpha$ galaxies in our field
as follows. Firstly, we assume a background Ly-$\alpha$ galaxy density
$\rho$ and an overdensity of Ly-$\alpha$ galaxies in our field $\omega$,
defined as the ratio of the mean Ly-$\alpha$ galaxy density in our field
divided by the background Ly-$\alpha$ galaxy density. Using
Poisson statistics, we then derive the probability of seeing two or more
galaxies in our field $P_f$, and the probability of seeing no galaxies in
Pritchet \& Hartwick's survey $P_p$. We then vary $\rho$ to
maximize the joint probability $P_j = P_f P_p$, for a given value of
$\omega$.

If our field is in no way special ($\omega = 1$), the peak joint probability
is $5 \times 10^{-5}$. We can therefore reject the null hypothesis that
Ly-$\alpha$ galaxies are not overdense in our field. For larger values
of $\omega$, the maximum joint probability increases, and occurs for smaller
$\rho$. To bring $P_j$ up to 5\% , we require $\omega > 15$; our field
must have an overdensity of a factor of at least 15 (95\% confidence).

An overdensity of $> 15$ in our 2$^{\prime}$ \ radius field, while consistent
with Francis \& Hewett's claim, is not in itself very surprising; 2$^{\prime}$
\
corresponds to a proper distance scale of $\sim 500$ kpc, and overdensities of
$> 15$ on these scales are common in the local universe. If however the
overdensity remains this high over the whole postulated supercluster,
$> 8 $ $^{\prime}$ \ in diameter, this would be surprising. The only object in
the local universe approaching such overdensities on these large scales
is the Great Wall. Heisler et al. (1989) and Wolfe (1993) argue that the
existence of structures this large and dense
at high redshifts is a challenge for gravitational structure formation
models such as CDM. The location of B2, C1 and C2 near the edge of our
field of view suggests that the galaxy overdensity does indeed continue
beyond the field we have imaged to date.

The main weakness with this analysis is with the control sample. Pritchet
\& Hartwick's sample is at the slightly lower redshift of $z \sim 1.9$.
If the number density of Ly-$\alpha$ galaxies was evolving strongly
between $z=2.38$ and $z \sim 1.9$ the overdensity in our field might be much
lower. De Propris et al. (1993) carried out a field search for
Ly-$\alpha$ emitting galaxies between redshifts 2--3 and discovered
none, but unfortunately their survey was insufficiently deep to detect
objects with fluxes comparable to B1 and B2. The many other unsuccessful
Ly-$\alpha$ galaxy searches, many at redshifts above 2.38 (eg.
Lowenthal et al. 1995, Djorgovski, Thompson \&
Smith 1993) however make us doubt
that the numbers of Ly-$\alpha$ galaxies are evolving strongly.
It is also worrying that Pritchet \& Hartwick were selecting their
candidates by eye; De Propris et al. (1993), using an almost
identical technique, were only finding $\sim 50$\% of simulated 10$\sigma$
galaxies. Clearly a quantitative analysis of deeper control fields at
$z\sim 2.4$ would help.

Assuming that our analysis is valid, and that the overdensity of galaxies
really does extend to Mpc scales, one possible explanation is that the
distribution of Ly-$\alpha$ emitting galaxies
is strongly decoupled from that of the underlying matter. If, for example,
Ly-$\alpha$ emitting galaxies only form at very high density peaks in the
underlying mass distribution, they will be far more strongly clustered than
matter (eg. Kaiser 1984). Brainerd \& Villumsen (1994)
showed that in n-body CDM simulations, dark matter halos cluster far more
strongly than the underlying mass at $z \sim 2$.
Evidence for such biassing comes from
the very different clustering amplitudes inferred for Ly-$\alpha$ forest
clouds and metal-line QSO absorption systems (Carswell \& Rees
1987, Heisler et al. 1989).

\subsection{The Nature of B1}

In recent years, a small number of Ly-$\alpha$ emitting galaxies
have been found near QSO absorption-line systems (Lowenthal et
al. 1991, M{\o}ller \& Warren 1993, Machetto et
al. 1993, Steidel, Sargent \& Dickinson  1991).  Most are
very compact; at best marginally resolved, unlike B1.

The morphology of B1 is however typical of high redshift
radio galaxies, which form the vast majority of known high redshift
galaxies ( McCarthy 1993). B1 resembles these galaxies in
many ways: in its elongated irregular morphology, in its Ly-$\alpha$
flux and equivalent width, in its line ratios, in its red continuum colors,
and in its red companion objects (Rigler et al. 1992). It
lies on the $K$-band Hubble diagram for powerful radio galaxies.
There are however two significant differences: B1's radio
flux is at least two orders of magnitude below that of radio galaxies
with comparable emission-line fluxes, and the velocity width of
Ly-$\alpha$ is about half that of a typical radio galaxy (though this
may be due to absorption). Nonetheless, the similarity between B1 and
many radio galaxies is striking.

B1 consists of an IR-bright unresolved core (rest-frame absolute $V$-band
magnitude of $\sim -24.4$), and an extended, asymmetric line-emitting tail
(Fig~3). As shown in \S~3.2, the core has a pronounced spectral break at
$\sim 1.3\mu$m observed-frame, giving it a rest-frame color of $U-V > 1$.
Our narrow-band H-$\alpha$ imaging shows that only $< 10$\% of the
$K^{\prime}$ flux comes from line emission, so we attempted to fit B1's
colors using synthetic galaxy spectra from Arimoto \& Yoshii
(1987), Arimoto, Yoshii \& Takahara (1992), Bruzual
\& Charlot (1993) and Rocca-Volmerange \& Guiderdoni (1988).
The magnitude of the spectral break can only be reproduced by stellar
populations with little ongoing or recent star formation; a population of
stars formed in an instantaneous burst and then allowed to passively age
for $> 0.6$ Gyr gives a good fit. If the age is $\sim 0.6$ Gyr, the $1.3 \mu$m
break would be due to Balmer absorption in A stars, whose light is expected to
dominate galaxies of this age. If older, the break would be the well known
$4000$\AA\ feature.
The stellar mass $M$ derived from the rest-frame
$V$-band light is somewhat model dependent, being somewhere in the
range $\sim 2 \times 10^{11} {\rm M}_{{\rm Sun}} < {\rm M} < 2 \times 10^{12}
{\rm M}_{{\rm Sun}}$. Models in which star formation is ongoing or recent
predict
$0.5 \mu$m fluxes significantly above our limits, due to a population of O
and B stars. If the tentative detection
of continuum flux at 5000 \AA \ in the red spectrum is correct, a small
number of blue stars or some non-thermal component would be required.
An age of $\sim 0.5$ Gyr at $z=2.38$ implies
(for $H_0 = 75 {\rm km\ s}^{-1}{\rm Mpc}^{-1}$) that formation occurred
at $z \sim 3.7$ ($q_o = 0.5$) or $z \sim 3.0$ ($q_0 = 0.1$). If the
age were 1Gyr, the epoch of formation would be $z \sim 5$
($q_0 = 0.5$)  or $z \sim 4$ ($q_0 = 0.1$).

What is the cause of the line emission? The Ly-$\alpha$ and H-$\alpha$
flux could be generated by star formation, at a rate of $\sim 100
{\rm \ M}_{{\rm Sun}}{\rm yr}^{-1}$ (Kennicutt 1983) without
overproducing blue continuum light (Charlot \& Fall 1993).
However the detection of C~IV strongly suggests that an AGN is present,
as hot stars produce few photons capable of ionizing carbon to
this state. The similarity between B1 and radio galaxies also suggests
that an AGN is present, albeit a radio-quiet one.

The Ly-$\alpha$ equivalent width of B1 is higher, and the Ly-$\alpha$
line narrower than those of typical QSOs (Francis et al. 1992,
Francis 1993). If
however the active nucleus were hidden from direct observation by
dust clouds, but its radiation could escape in other directions, we
would only see scattered light and line emission from
photoionized gas far from the nucleus, which might have much higher
equivalent widths and lower velocity widths. This is now known to
be the explanation of strong narrow lines of many Seyfert 2 galaxies
(eg. Antonucci \& Miller 1985), and is widely invoked to
explain the line emission of radio galaxies.

Many high redshift QSOs and radio galaxies show extended Ly-$\alpha$
emission with fluxes in excess of B1's, originating on scales of
up to 100 kpc (eg. Heckman et al. 1991). An AGN with
an absolute $B$ magnitude of $-23$ produces enough ionizing
photons to generate the observed Ly-$\alpha$ and H-$\alpha$ fluxes.
If the Ly-$\alpha$ flux were produced by photoionization from an AGN
concealed in the IR-bright core, the asymmetry of the emission would
have to be caused by a highly asymmetric, lumpy gas distribution. Our
upper limit on the radio flux of B1 is not inconsistent with the typical
radio fluxes of the least radio-loud QSOs (Kellerman et al. 1989).

B1's Ly-$\alpha$ emission peaks $\sim 490 {\rm \ km\ s}^{-1}$ to the
red of its C~IV emission (Fig~5). We suggest that this offset is caused by
Ly-$\alpha$ absorption of the blue part of the Ly-$\alpha$ line.
If the absorbing neutral hydrogen is dust-free, it cannot be physically
associated with the emission-line region, as it would scatter roughly
as many photons into the line of sight as out of it (unless the geometry
was very special). If the absorber
is indeed a foreground screen of neutral hydrogen, its redshift and
column density would have to be similar to those of the central component
of the QSO absorption-line. The Ly-$\alpha$ emission would need to
be intrinsically red asymmetric, and to have an intrinsic velocity width
of $\sim 1000 {\rm \ km\ s}^{-1}$.

If, alternatively, the absorption occurs in neutral hydrogen mixed
with the emission-line gas, dust is needed to absorb the resonantly
scattered Ly-$\alpha$ photons. The red wing of Ly-$\alpha$ might escape
because the high velocity gas is dust-free, or because the optical
depth of the high velocity gas is low.

We conclude that B1 is probably a large early type galaxy, containing
a hidden radio-quiet QSO, which is photoionizing the lumpy surrounding
gas.

\subsection{Evidence for Dust}

The observed Ly-$\alpha$ to H-$\alpha$ ratio of B1 is $\sim 0.8$; well
below the case B recombination value of 8. This strongly suggests that
dust is present. The Ly-$\alpha$/H-$\alpha$ ratio is low both in the
Ly-$\alpha$ knot and the tail, so the dust must extend over at least
50 kpc.

If all the dust is in a foreground screen, an extinction $E(B-V) \sim 1$
is required to explain the observed ratio. Much smaller amounts of dust will
suffice if it is mixed in with the gas, as the optical depth of the
gas to resonant scattering of Ly-$\alpha$ is likely to be large, giving the
Ly-$\alpha$ photons many chances to be absorbed.

The presence of extended dust in B1, associated with the QSO absorption-line
system, supports the contention of Fall \& Pei (1993) that
high column density Ly-$\alpha$ QSO absorption-line systems contain dust, and
therefore can obscure background QSOs. Given the steepness of the
QSO luminosity function, even very small quantities of dust in
absorption-line systems will seriously bias absorption-line statistics
and QSO number counts (eg. Webster et al. 1995).

\subsection{The Environment of B1}

The region around B1 and the QSO is an extraordinary one. Within
a radius of $\sim 10^{\prime \prime}$ we have three Ly-$\alpha$ absorption
components,
one Ly-$\alpha$ emitting galaxy and possibly three less active
galaxies, C3, C4 and C5. If they are physically associated,
they all lie within $\sim 100$ kpc of each other.

If C3, C4 and C5 really are inactive galaxies at $z=2.38$, then together
with B1 they form a high redshift analogue of a compact group
(Hickson, Kindl \& Huchra 1989). Many more galaxies
could be present but below our detection threshold. The extended
irregular emission of B1 suggests that this group of galaxies is embedded
in a lumpy cloud of dust and gas.

The absorption in the spectrum of QSO 2139$-$4434 also suggests the presence
of a lumpy gas cloud. It is interesting that the column density of
the main QSO absorption component is similar to that needed to absorb the
blue wing of B1's Ly-$\alpha$ emission (\S~4.3). If the QSO absorption
components really are associated with the gas cloud embedding B1, C3, C4 and
C5, we can estimate the mass of this gas cloud: a
gas cloud of column density $N_H \sim 10^{19} {\rm cm}^{-2}$ and radius
50 kpc will have a neutral hydrogen mass of $\sim 2 \times 10^9 M_{{\rm Sun}}$,
ie. $\sim 5$\% of the stellar mass of B1.

On a yet more speculative note, if all three absorption-line components
are associated with the galaxy group, we can combine their redshifts with
that of B1 to estimate a velocity dispersion for the gas cloud: $\sim 500
{\rm \ km\ s}^{-1}$. This calculation should be regarded with great
caution; quite apart from the small number statistics, the velocities of
the three absorption-line components could
be dominated by the gravity of a compact galaxy hiding in the point spread
function of the QSO. Alternatively, they may lie in different parts of
Francis \& Hewett's supercluster; if the absorption-line supercluster is
roughly spherical and expanding with the Hubble flow, it will have
an expected radial extent of $\sim 1000 {\rm \ km\ s}^{-1}$, greater than
the velocity differences between the different absorption-line components.
If however $500 {\rm \ km\  s}^{-1}$ is a reasonable estimate of the
velocity dispersion of the group, and this velocity dispersion is dominated
by gravitational motions and not by gas dynamics, the virial mass of
the group is $\sim 10^{12} {\rm M}_{{\rm Sun}}$ and the crossing-time is
$\sim 10^8$ years.

Irrespective of the velocity dispersion of the group, if B1 and its
neighbors are gravitationally bound, their merger timescale will
be $\ll 1$ Gyr (but see Governato, Bhatia \& Chincarini 1991).
We therefore speculate that by redshift zero, B1 will have accreted
C3, C4, C5 and much of its gaseous halo, forming a massive elliptical galaxy,
perhaps even a cD galaxy.

To summarize, if B1, the color-selected galaxies and the QSO absorption line
clouds really are physically associated, they form an extremely gas rich
compact group, which may be a giant elliptical galaxy in the act of
bottom-up formation.

\section{Conclusions}

We have detected and confirmed two Ly-$\alpha$ emitting galaxies at
$z=2.38$, associated with a cluster of Ly-$\alpha$ QSO absorption
lines. One of
these galaxies, B1, lies 22$^{\prime \prime}$\ from a background QSO, and is
surrounded
by a group of very red objects, some of which may be other
galaxies at the same redshift. The detection of two Ly-$\alpha$ emitting
galaxies in the small area surveyed supports Francis \& Hewett's claimed
supercluster, though final proof will require the mapping of the rest of
the postulated supercluster.

B1 is a massive early-type galaxy, at least 500 Myr old, and probably
contains a concealed radio-quiet AGN, which is photoionizing an irregular
cloud of gas surrounding B1. Together with its neighboring red galaxies, it
may be a giant elliptical galaxy in the act of bottom-up formation.

We have detected H-$\alpha$ emission from B1 at a level indicating that
dust is present. This supports Hu et
al. (1993) and Bunker et al.'s (1995) contention that H-$\alpha$
searches may be a powerful tool for finding high redshift galaxies. The dust
is extended over at least 50 kpc, supporting Fall \& Pei's (1993)
contention that dust obscuration is important in QSO absorption lines.

The pronounced $1.3\mu$ break in the spectrum of B1 suggests an additional
technique for finding high redshift galaxies. With optical and near-IR
photometry, it may be possible to select galaxies by the presence of a
redshifted Balmer or $4000$\AA\ break. The technique is similar to
Steidel \& Hamilton's (1992) broad-band selection of galaxies
showing Ly-limit breaks, but is sensitive to lower redshifts and
older galaxies.

\bigskip

We wish to thank Ken Freeman, Dick Hunstead, Bruce Peterson and Rachel
Webster for helpful discussions. Dick Hunstead and Taisheng Ye kindly made
their Molonglo radio observations available to us.
PJF is supported by an ARC grant, and acknowledges travel support from
ANSTO. MM acknowledges the support of a University of Melbourne
Fellowship for Women with Career Interruptions. AJB acknowledges a
UK Particle Physics and Astronomy Research Council studentship. YY
acknowledges financial support of the Yamada Science Foundation for
transport of the PICNIC camera and IR team. Astrometry was
done using coordinates from the COSMOS database, maintained on-line by the
AAO.

\bigskip

\begin{table}

\title{Table 1}

\begin{tabular}{lccc}
Candidate & Coordinates & Ly-$\alpha$ Flux & Ly-$\alpha$ Equivalent
Width   \\
& (B1950) & (erg\ ${\rm cm}^{-2}{\rm s}^{-1}$) & ($3 \sigma$, \AA\
observed-frame)  \\
\hline
B1 & 21:39:16.32 - 44:34:12.9 & $8 \pm 1 \times 10^{-16}$ & $> 270$  \\
B2 & 21:39:18.50 - 44:34:44.9 & $3 \pm 0.3 \times 10^{-16}$ & $> 160$  \\
B3 & 21:39:16.82 - 44:34:40.7 & $0.8 \pm 0.3 \times 10^{-16}$ & --- \\
2139-4434 & 21:39:14.62 - 44:34:00.5 & --- \\
\end{tabular}

\end{table}

\section{References}

\vspace{2mm} \noindent
Arimoto, N., \& Yoshii, Y. 1987, A \& A, 173, 23

\vspace{2mm} \noindent
Arimoto, N., Yoshii, Y., \& Takahara, F. 1992, A \& A, 253, 21

\vspace{2mm} \noindent
Antonucci, R. R. J., \& Miller, J. S. 1985, ApJ, 297, 621

\vspace{2mm} \noindent
Baron, E., \& White, S. D. M. 1987, ApJ, 322, 585

\vspace{2mm} \noindent
Brainerd, T. G., \& Villumsen, J. V. 1994, ApJ, 431, 477

\vspace{2mm} \noindent
Bruzual A. G., \& Charlot, S. 1993 ApJ, 405, 538

\vspace{2mm} \noindent
Bunker, A. J., Warren, S. J., Hewett, P. C., \&
Clements, D. L. 1995, MNRAS, 273, 513

\vspace{2mm} \noindent
Butcher, H., \& Oemler, A. 1978 ApJ, 219, 18

\vspace{2mm} \noindent
Carswell, R. F., \& Rees, M. J. 1987, MNRAS, 224, 13p

\vspace{2mm} \noindent
Colless, M., Ellis, R. S., Taylor, K., \& Hook, R. N. 1990,
MNRAS, 244, 408

\vspace{2mm} \noindent
Charlot, A., \& Fall, S. M. 1993, in First Light in the
Universe, Stars or QSOs?  ed. B. Rocca-Volmeragne et al.
(Gif-sur-Yvette: Editions Fronti\'eres), 341

\vspace{2mm} \noindent
De Propris, R., Pritchet, C. J., Hartwick, F. D. A., \&
Hickson, P. 1993, AJ, 105, 1243

\vspace{2mm} \noindent
Dekker, H., D'Odorico, S., Kotzlowski, K., Lizon, J.-L.,
Longinotti, A., Nees, W., \& De Lapparent-Gurriet, V. 1991, ESO
messenger, 63, 73

\vspace{2mm} \noindent
Djorgovski, S., Spinrad, H., McCarthy, P., \& Strauss, M. A.
1985, ApJl, 299, L1

\vspace{2mm} \noindent
Djorgovski, S., Thompson, D., \& Smith, J. D. 1993, in First
Light in the Universe, Stars or QSOs? ed. B. Rocca-Volmerange et al.
(Gif-sur-Yvette: Editions Fronti\'eres), 67

\vspace{2mm} \noindent
Dressler, A., Oemler, A., Gunn, J. E., \& Butcher, H. 1993,
ApJl, 404, L45

\vspace{2mm} \noindent
Eckert, W., Hofstadt, D., \& Melnick, J. 1989, ESO messenger,
57, 66

\vspace{2mm} \noindent
Edge, A. C., Stewart, G. C., Fabian, A. C., \& Arnaud, K. A.
1990, MNRAS, 245, 559

\vspace{2mm} \noindent
Efstathiou, G., Bernstein, G., Katz, N., Tyson, J. A., \&
Guhathakurta, P. 1991, ApJl, 380, L47

\vspace{2mm} \noindent
Evrard, A. E., \& Charlot, S. 1994, ApJl , 424, L13

\vspace{2mm} \noindent
Foltz, C. B., Hewett, P. C., Chaffee, F. H., \& Hogan, C. J.
1993, AJ, 105, 22

\vspace{2mm} \noindent
Fall, S. M., \& Pei, Y. C. 1993, ApJ, 402, 479

\vspace{2mm} \noindent
Francis, P. J. 1993, ApJ, 405, 119

\vspace{2mm} \noindent
Francis, P. J., \& Hewett, P. C. 1993, AJ, 105, 1633

\vspace{2mm} \noindent
Francis, P. J., Hewett, P. C., Foltz, C. B., \&  Chaffee, F. H.,
1992, ApJ, 398, 476

\vspace{2mm} \noindent
Gebhardt, K., Pryor, C., Williams, T. B., \& Hesser, J. E. 1994,
AJ, 107, 2067

\vspace{2mm} \noindent
Giavalisco, M., Steidel, C. C., \& Szalay, A. S. 1994, ApJl, 425, L5

\vspace{2mm} \noindent
Glazebrook, K., Ellis, R., Colless, M., Allington-Smith, J.,
\& Tanvir, N. 1995, MNRAS, 273, 157

\vspace{2mm} \noindent
Governato, F., Bhatia, P., \& Chincarini, G. 1991, ApJl, 371, L15

\vspace{2mm} \noindent
Heckman, T. M., Lehnert, M. D., van Breugel, W., \&
Miley, G. K. 1991, ApJ, 370, 78

\vspace{2mm} \noindent
Heisler, J., Hogan, C. J., \& White, S. D. M. 1989, ApJ, 347, 52

\vspace{2mm} \noindent
Hickson, P., Kindl, E., \& Huchra, J. P., 1989, ApJ, 331, 64

\vspace{2mm} \noindent
Hu, E. M., Songaila, A., Cowie, L. L., \& Hodepp, K. -W. 1993,
ApJl, 419, L13

\vspace{2mm} \noindent
Jakobsen, P., \& Perryman, M. 1992, ApJ, 392, 432

\vspace{2mm} \noindent
Kaiser, N. 1984, ApJl, 284, L9

\vspace{2mm} \noindent
Kellerman, K., Sramek, R., Schmidt, M., Shaffer, D., \& Green, R.
1989, AJ, 98, 1195

\vspace{2mm} \noindent
Kennicutt, R. C. 1983, ApJ, 272, 54

\vspace{2mm} \noindent
Kobayashi, Y., Fang, G., Minezaki, T., Waseda, K., \& Nakamura, K.
1994 in SPIE proceedings ``Instrumentation in Astronomy 8'', 2198, 603

\vspace{2mm} \noindent
Leggett, S. K. 1992, ApJs, 82, 351

\vspace{2mm} \noindent
Lowenthal, J. D., Hogan, C. J., Green, R. F., Caulet, A.,
Woodgate, B. E., Brown, L., \& Foltz, C. B. 1991, ApJl, 377, L73

\vspace{2mm} \noindent
Lowenthal, J. D., Hogan, C. J., Green, R. F., Woodgate, B. E.,
Caulet, A., Brown, L., \& Bechtold, J. 1995, ApJ, in press

\vspace{2mm} \noindent
Machetto, F., Lipari, S., Giavalisco, M., Turnshek, D. A.,
\& Sparkes, W. B. 1993, ApJ, 404, 511

\vspace{2mm} \noindent
McCarthy, P. J. 1993, Ann Rev Ast. Ast., 31, 639

\vspace{2mm} \noindent
M{\o}ller, P., 1995, A \& A, in press

\vspace{2mm} \noindent
M{\o}ller, P., \& Warren, S. J., 1993, A \& A, 270, 43

\vspace{2mm} \noindent
Pritchet, C. J., \& Hartwick, F. D. A. 1990, ApJl, 355, L11

\vspace{2mm} \noindent
Rees, M. J. 1988, MNRAS, 231, 91p

\vspace{2mm} \noindent
 Rigler, M. A., Lilly, S. J., Stockton, A., Hammer, F., \&
LeF\`evre, O., 1992, ApJ, 385, 61

\vspace{2mm} \noindent
Rocca-Volmerange, B., \& Guiderdoni, B. 1988, A \& A Supp, 75, 93

\vspace{2mm} \noindent
Rodrigues-Williams, L. L., \& Hogan, C. J. 1994, AJ, 107, 451

\vspace{2mm} \noindent
Steidel, C. C., \& Hamilton, D. 1992, AJ, 104, 941

\vspace{2mm} \noindent
Steidel, C. C., Sargent, W. L. W., \& Dickinson, M. 1991, AJ, 101,
1187

\vspace{2mm} \noindent
Webster, R. L., Hewett, P. C., Harding, M. E., \& Wegner, G. A.
1988, Nature, 336, 358

\vspace{2mm} \noindent
Webster, R. L., Francis, P. J., Peterson, B. A., Drinkwater, M. J.,
\& Masci, F. J. 1995, Nature, 375, 469

\vspace{2mm} \noindent
Williger, G. M., Hazard, C., Baldwin, J. A., \& McMahon, R. G.
1995 ApJ, submitted

\vspace{2mm} \noindent
Wolfe, A. M. 1993, ApJ, 402, 411

\newpage

\begin{figure}
% fig 1
\caption{Ly-$\alpha$ On-band (left) and off-band (right) images of
the field around QSO 2139-4434. Images have been lightly smoothed
by a Gaussian of $\sigma=0.6^{\prime \prime}$, axes
are in B1950 coordinates. See Fig~2 for the names of the objects.}
\end{figure}

\begin{figure}
% Fig 2

\caption{Identification chart for the images. The orientation and field of
view of this chart is the same as that of figures 1 and 6. Solid triangles
are Ly-$\alpha$ emission candidates; squares are objects with large
$I-K^{\prime}$ colors; ie. candidate cluster members (\S~3.2). Circles are
objects red in $B-K^{\prime}$ but not in $I-K^{\prime}$; ie. candidate
foreground galaxies. Other objects are marked with crosses and are probably
foreground stars and galaxies.}

\end{figure}

\begin{figure}
% Fig 3

\caption{Close-up of B1. The $K^{\prime}$ image is shown in greyscales, and
the narrow-band Ly-$\alpha$ image (smoothed with a Gaussian beam of $\sigma
= 0.56^{\prime \prime}$ ) is shown by contours. Contours are equally spaced;
the
zero level contour is dotted. The internal astrometry of the $K^{\prime}$
and Ly-$\alpha$ images is good to $\sim 0.15^{\prime \prime}$, based on the rms
fit of the positions of all the images; the misalignment of the object
in the top right is also real.}
\end{figure}

\begin{figure}
% Fig 4
\caption{NTT Spectra of B1 and B2 and the QSO 2139$-$4434, showing
the Ly-$\alpha$ lines. A higher resolution spectrum of the QSO can be
found in Francis \& Hewett (1993).}
\end{figure}

\begin{figure}
% Fig 5
\caption{C IV and Ly-$\alpha$ emission from B1 (NTT spectra).
The Ly-$\alpha$ absorption
line in the spectrum of QSO 2139$-$4434 is shown for comparison.}
\end{figure}

\begin{figure}
% Fig 6

\caption{$I$-band and Siding Spring 2.3-m $K^{\prime}$-band (right) images.
The white patches above and below the brightest sources are artifacts of
the data reduction.}

\end{figure}

\begin{figure}
% Fig 7
\caption{Spectral energy distributions for a number of candidate galaxies.
The distributions marked (a) are for two representative red objects showing
a break between $B$ and $I$; ie. candidate foreground galaxies. The different
objects have displaced vertical scales for clarity. The data points are for
standard $B$, $I$, $J$, $H$ and $K^{\prime}$ bands except for B1, where a
more stringent limit on the blue flux is placed at 5000\AA\ from our spectra.}

\end{figure}

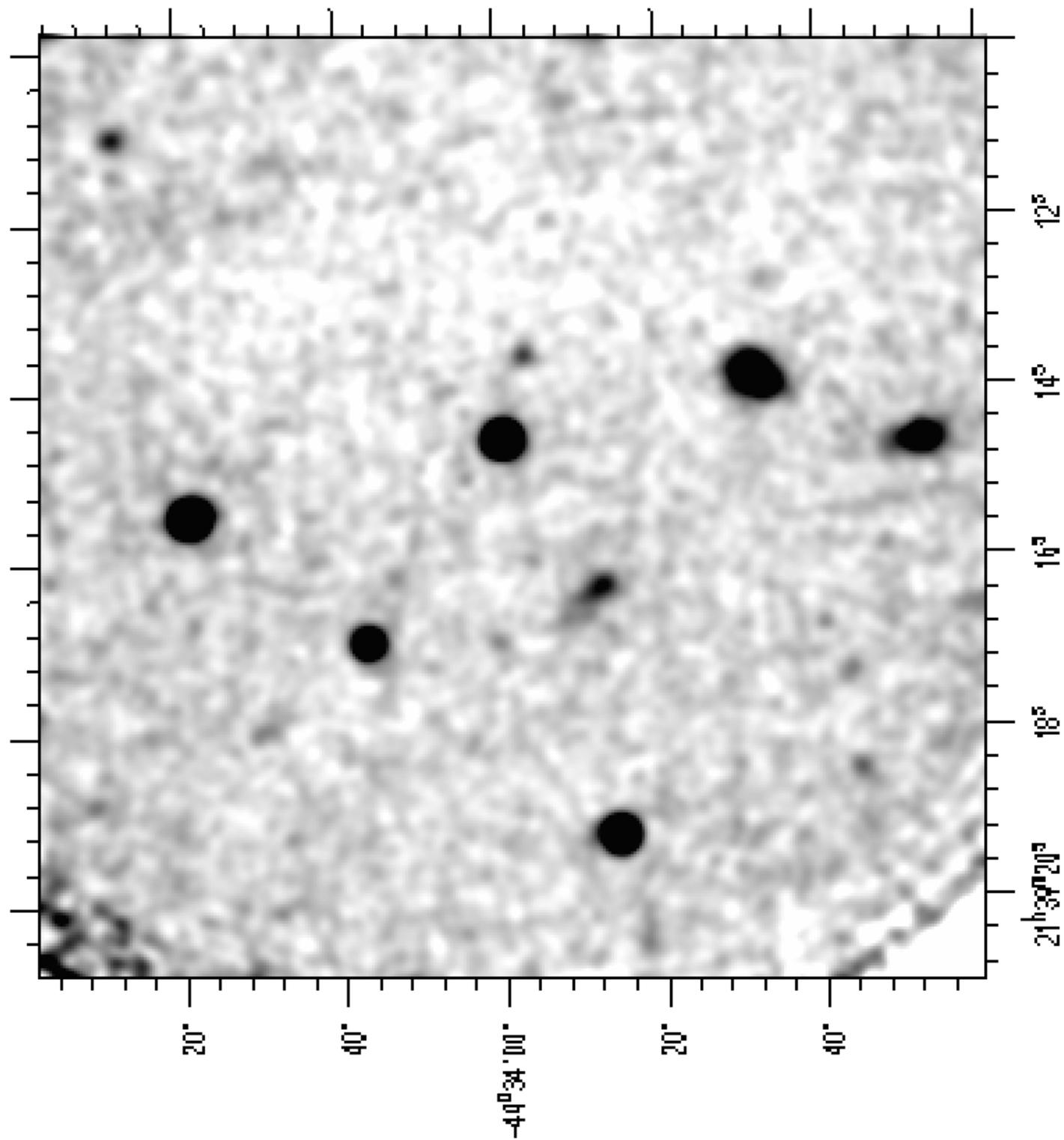
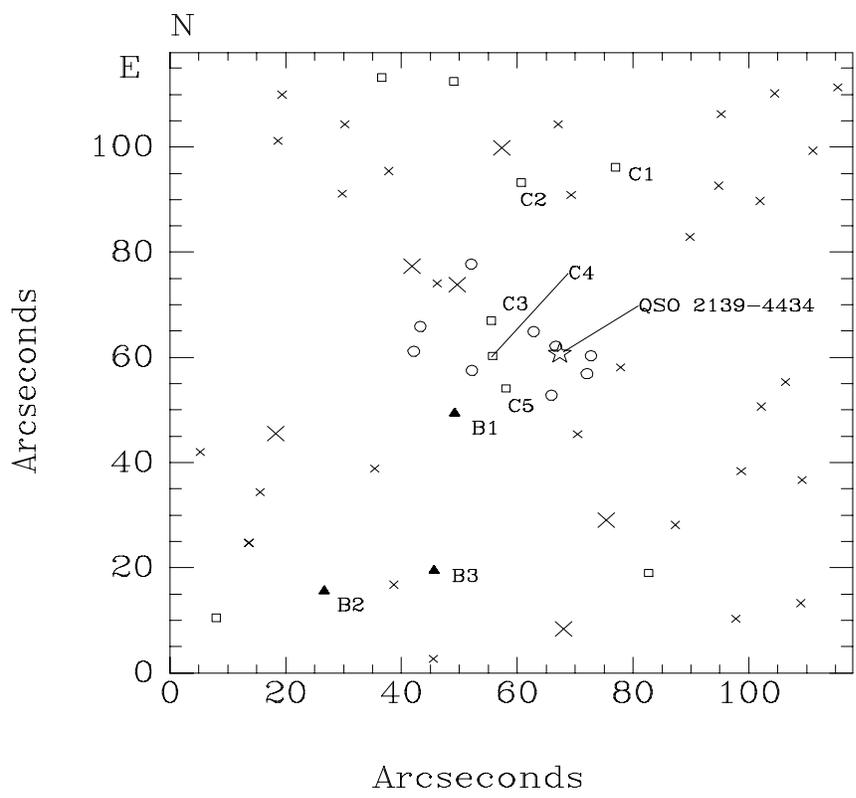
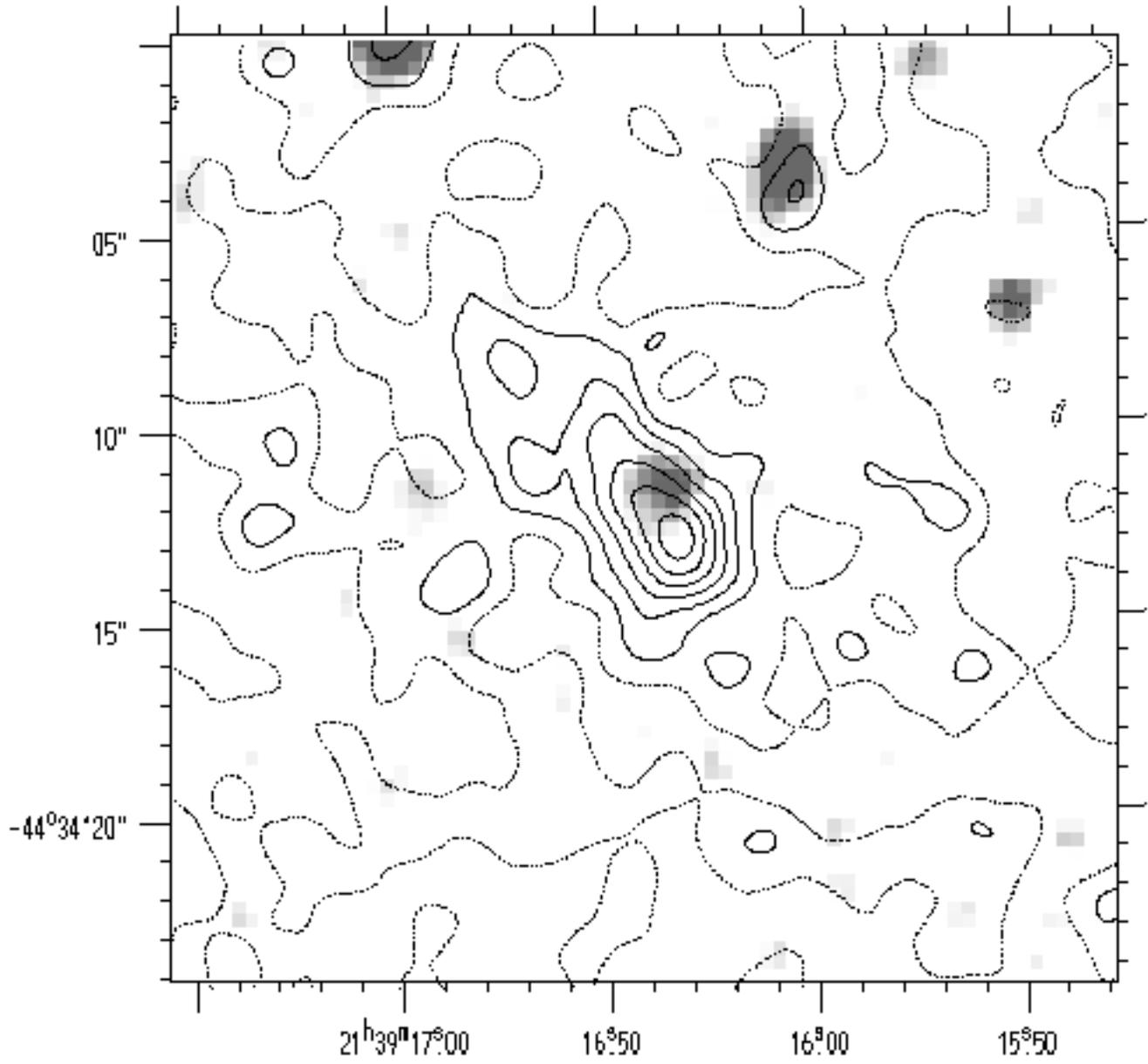
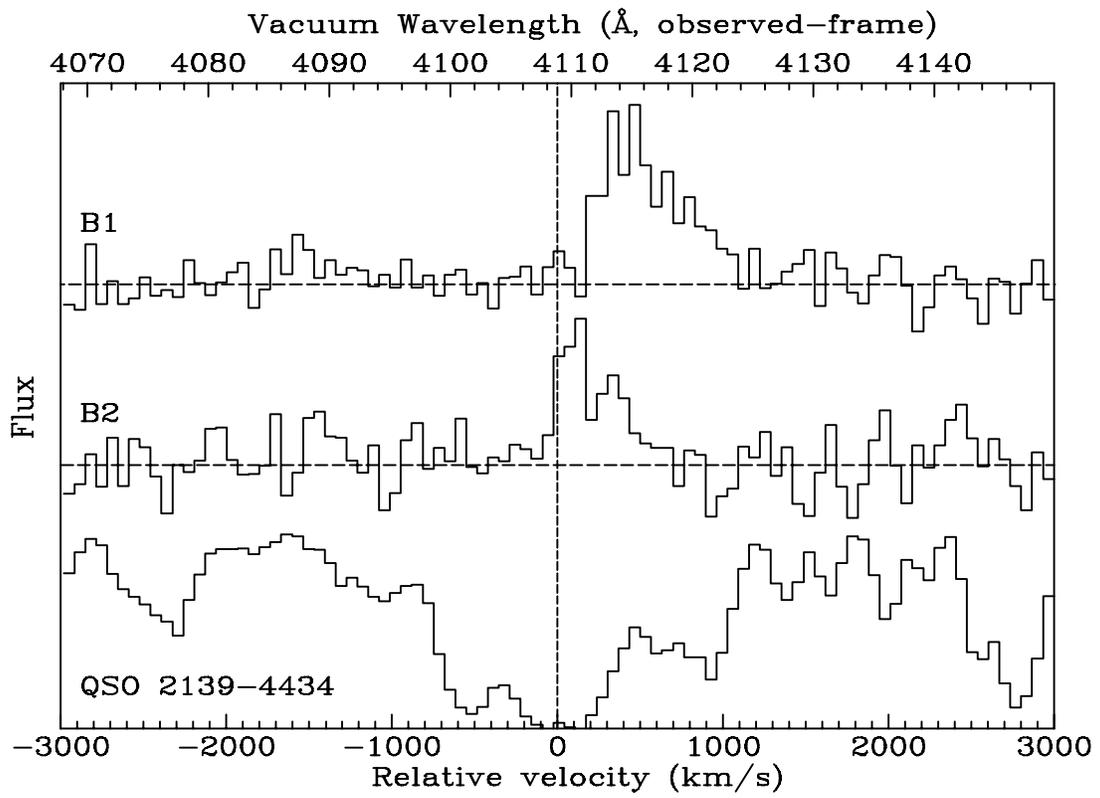
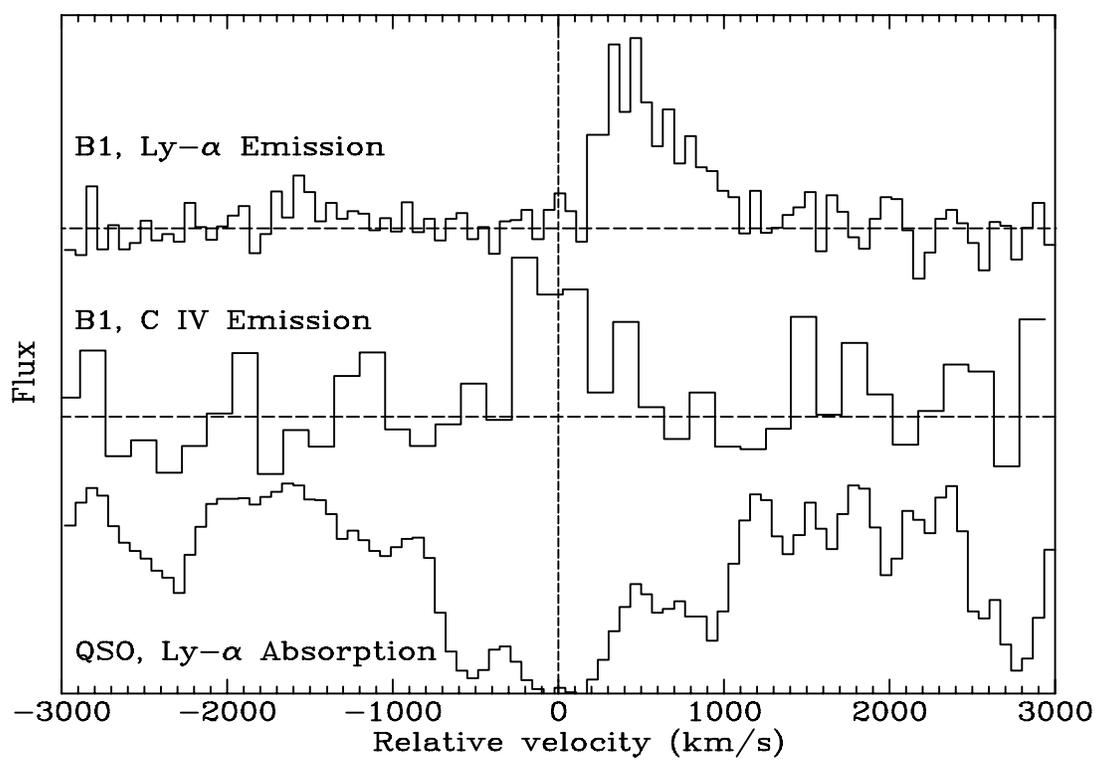
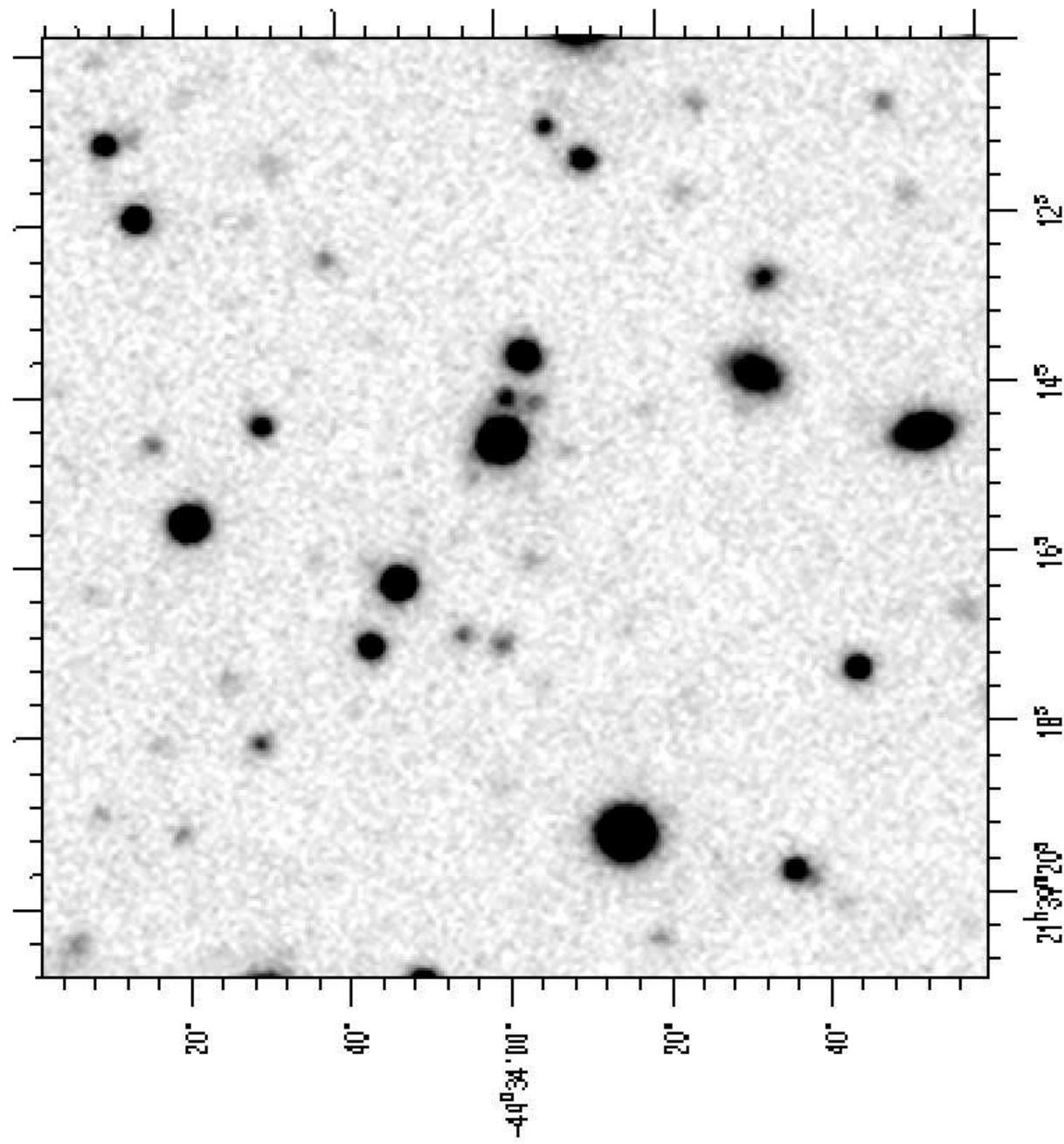
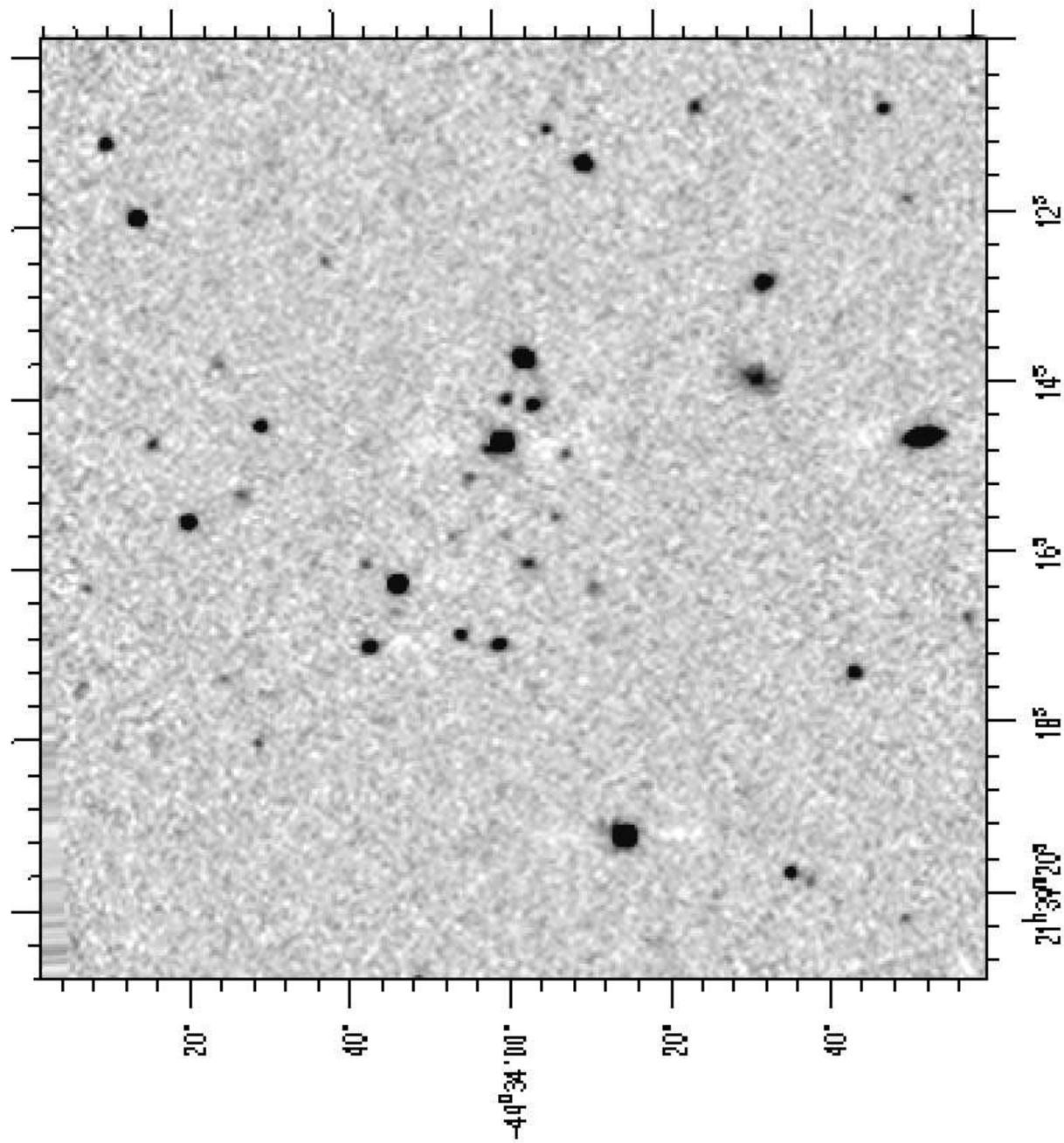
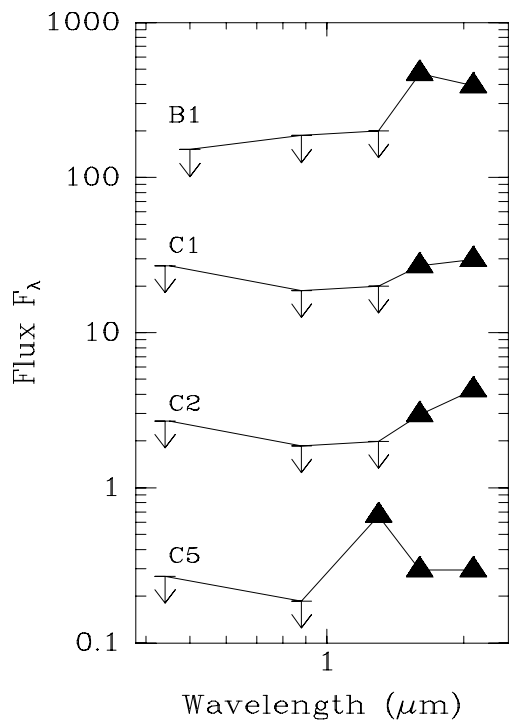
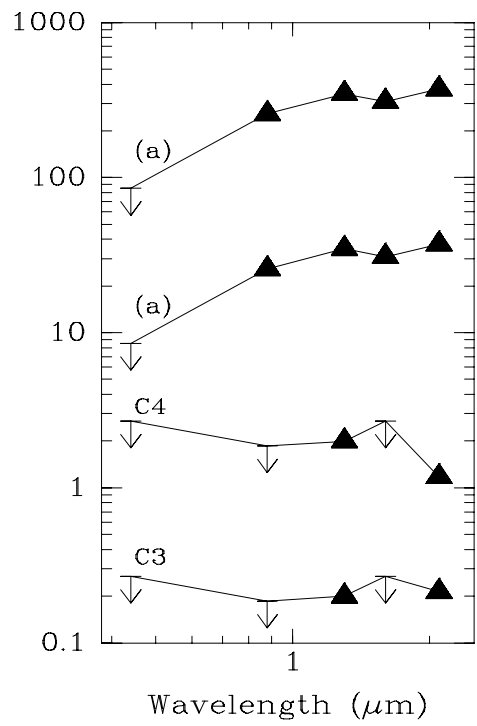
\end{document}